\documentclass[prd,twoside,twocolumn,showpacs,aps,longbibliography]{revtex4-2}
\usepackage{amsmath}
\usepackage{bbm}
\usepackage{amssymb}
\usepackage{amsfonts}

\renewcommand{\vec}{\textbf}
\newcommand{\ket}[1]{|#1\rangle}
\newcommand{\bra}[1]{\langle#1|}
\newcommand{\bracket}[2]{\langle#1|#2\rangle}
\DeclareMathOperator{\tr}{Tr}
\DeclareMathOperator{\RLGMT}{\mathsf{Realignment}}
\usepackage{graphicx}
\usepackage{color}

\begin{document}

\title{Is bound entanglement Lorentz invariant?}

\author{Pawe{\l}{} Caban}
\email{Pawel.Caban@uni.lodz.pl}
\affiliation{Department of Theoretical Physics,
	Faculty of Physics and Applied Informatics, University of Lodz
Pomorska 149/153, 90-236 {\L}{\'o}d{\'z}, Poland}

\author{Beatrix C. Hiesmayr}
\email{Beatrix.Hiesmayr@univie.ac.at}
\affiliation{University of Vienna, Faculty of Physics, Boltzmanngasse 5, 1090 Vienna, Austria}

\begin{abstract}
Bound entanglement, in contrast to free entanglement, cannot be distilled into
maximally entangled states by two
local observers applying measurements and utilizing classical communication.
In this paper we ask whether a relativistic observer classifies states according
to being separable, bound or free entangled in the same manner as an unboosted observer.
Surprisingly, this turns out not to be the case. And that even if the system
in a given inertial frame of reference is separable with respect to the
partition momenta versus spins. In detail, we show that if the spin state is initially bound entangled, some boosted
observers observe their spin states to be either bound entangled, separable or free entangled. This also explains why a general measure of the entanglement property is difficult to find.
\end{abstract}
\maketitle

\section{Introduction}

Detecting entanglement, even given the full information of the physical state,
namely the density matrix is a NP-hard problem~\cite{Gurvits2003},
because of the existence
of bound or PPT (positive partial transposition) entangled states~\cite{HHH_1998-bound}.
Those states cannot be detected by taking
the partial transpose in one subsystem and finding at least one eigenvalue negative,
in which case we are dealing with \textit{free} entanglement.
This mathematical property of the density matrix has crucial physical implementation,
i.e. an ensemble of free entangled states can always be distilled to maximally
entangled states by local operations and classical communication (LOCC),
in strong contrast to bound entangled states. Those states can be generated by
maximally entangled states, Bell states, but this entanglement is then bounded,
i.e. cannot be distilled. This aspect of entanglement gave raise to a lot of speculations
why Nature provides us with that kind of entanglement and for what it could be
useful~\cite{HHHO2005,VB2014}.
Entanglement and other aspects of quantum information theory in the relativistic
setting were discussed in many papers, see, e.g., \cite{Czachor1997_1,PST2002,CR2005,Caban_2007_Photons_PRA,CRW2009,FBHH2010,PVD2011,TA_2013,SV2013,OH2021-single-particle-entangl,Lee2022} and references therein.
However, up to our best knowledge, the behavior of bound entanglement under
Lorentz boosts was not analyzed up to now. One of the reasons is that bound
entanglement is difficult to detect.
However, recent works~\cite{Hiesmayr2021,PH2022,PH2022_Qutrits-Ququarts}
have shown some new insights on the structure
of bound entangled states in the Hilbert
space for the lowest dimensional cases of two qutrits or two ququarts.
We use those results in our present work.

\begin{figure}[h!]
\includegraphics[width=0.8\columnwidth,keepaspectratio=true]{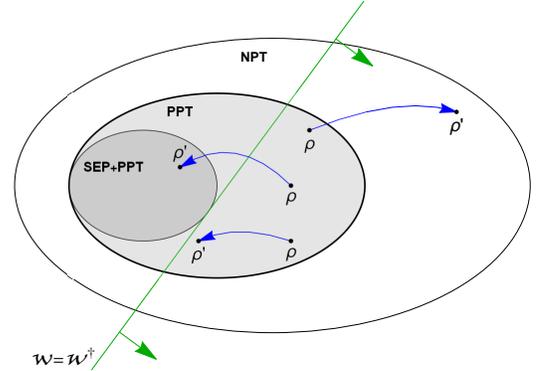}
\caption{This visualizes the state space and the effect of a relativistic boost on
bound entangled state.
An entanglement witness is visualized as a line separating some entangled states
from separable states.}
\label{fig-SEP-PPT-NPT}
\end{figure}

In this paper we analyze how bound entanglement changes under Lorentz boosts.
To this end we consider a system of two massive spin-1 particles.
In a one inertial frame of reference this system is prepared in a state that
is separable with respect to the partition momenta versus spins and the spin
part of this state is bound entangled.
We show that there exist such states and boosts that the boosted state is
also bound entangled or separable or even free entangled. This is also visualized in Fig.~\ref{fig-SEP-PPT-NPT}.

\section{Two-qutrit bound entangled states}

Entanglement is a genuine quantum feature of multipartite systems.
However, even in a bipartite case complete characterization of entanglement
can be given only for systems with dimensions $2\otimes2$ and $2\otimes3$.
In these dimensions one can fully characterize entanglement with the help of
the Peres--Horodecki Positive Partial Transpose (PPT) criterion
\cite{Peres1996_PhysRevLett.77.1413,HHH1996_PPT-criterion}.
This criterion says that if $\rho_{AB}^{T_B}$
(where the superscript $T_B$ denotes partial transposition with respect
to the system $B$)
is not positive semidefinite then the state $\rho_{AB}$ is entangled.
For $2\otimes2$ and $2\otimes3$ dimensional systems also the inverse is true.
However, for higher dimensional systems there exist entangled states
$\rho_{AB}$ such that $\rho_{AB}^{T_B}$ is positive semidefinite
\cite{HHH_1998-bound}.
Such states are called bound entangled, in opposition to free entangled
states for which $\rho_{AB}^{T_B}$ possesses at least one negative eigenvalue.
These names---free and bound entanglement---come from the fact that free
entanglement can be distilled while it is impossible for bound entanglement.

The detection of bound entanglement is not an easy task---given method can
certify entanglement of a certain family of states while it can be useless for other~\cite{Hiesmayr2021}.
One of the most useful and easiest in application methods is the realignment
or computable cross-norm criterion
\cite{ChenWu2003-realignment,Rudolph2003_CCR_PhysRevA.67.032312}, on which
we focus firstly. In the last Section~\ref{MUBWitness} we show how those
properties of states can be transferred via so called entanglement witnesses,
i.e. hermitian observables, to an experimental realization.
Let us denote
\begin{equation}\label{realignment}
	\RLGMT(\rho_{AB})= \log_2 \Big(
	\sum_i \sigma_i(\tilde{\rho}_{AB})
	\Big),
\end{equation}
where $\sum_i \sigma_i(\tilde{\rho}_{AB})$ is the sum of all the singular values
of the realigned matrix $\tilde{\rho}_{AB}$,
where $[\tilde{\rho}_{AB}]_{ij,ab}=[\rho_{AB}]_{ia,jb}$.
The realignment criterion says that if $\RLGMT(\rho_{AB})>0$ then the state
$\rho_{AB}$ is entangled. The above statement cannot be inverted---there exist
entangled states $\rho$ for which $\RLGMT(\rho)<0$.
Moreover, this criterion can detect bound entanglement of certain classes of states.

Bound entanglement can be also revealed by other methods like the quasispin criterion~\cite{Hiesmayr2009-Quasispin} or
with the help of entanglement witnesses (here the so called
MUB witnesses~\cite{SHBAH2012-MUB,HMBRBCh2021_MUB,BBChHM2021_MUB}
are  useful), which we discuss in Section~\ref{MUBWitness}.

In this paper we use bound entangled states
from the so called magic simplex for which recently a classification into separable,
bound entangled and free entangled states was possible with a success probability
of $95\%$ for qutrits~\cite{PH2022} and for ququarts
of $75\%$~\cite{PH2022_Qutrits-Ququarts}.
In the $3\otimes3$ dimensional case this simplex has the form~\cite{BHN2006,BHN2008}
\begin{equation}
	\mathcal{M}_3 \equiv \Big\{ \rho = \sum_{k,l=0}^{2} c_{k,l} P_{k,l}~ |~
	\sum_{k,l=0}^{2} c_{k,l} = 1, c_{k,l} \geq 0  \Big\},
	\label{MagicSimplex}
\end{equation}
where $P_{k,l}=\ket{\Omega_{k,l}}\bra{\Omega_{k,l}}$ and the Bell states
$\ket{\Omega_{k,l}}$ can be generated from
$\ket{\Omega_{0,0}}=\frac{1}{\sqrt{3}}(\ket{00}+\ket{11}+\ket{22})$ via the
relation $\ket{\Omega_{k,l}}= W_{k,l}\otimes\mathbbm{1}_3\; \ket{\Omega_{0,0}}$. In the last equation $W_{k,l}$ are the unitary Weyl operators
$ W_{k,l} \equiv \sum_{j=0}^{2} \omega^{j \cdot k}\; \ket{j} \bra{j+l}$
with $\omega = e^{\frac{2 \pi i}{3}}$ being the root of unity.

In particular we will use an
interesting one parameter state~\cite{Hiesmayr2021}
\begin{equation}
\rho_b(x) = \sum_{k,l=0}^{2} d_{k,l} \ket{\Omega_{k,l}}
\label{rho-bound}
\end{equation}
with
\begin{equation}
\begin{pmatrix}
d_{0,0}& d_{0,1}& d_{0,2}\\
d_{1,0}& d_{1,1}& d_{1,2}\\
d_{2,0}& d_{2,1}& d_{2,2}
\end{pmatrix}
=
\begin{pmatrix}
2x& 0& \frac{1}{3}-x\\
0& x& \frac{1}{3}-x\\
0& 0&\frac{1}{3}-x
\end{pmatrix}
\end{equation}
and $0\le x \le \tfrac{1}{3}$.
This state is PPT for $x\in[0,\frac{2}{15}]$, however, entangled for all values
$x$ except $x=0$. This means that for $x\in\{0,\frac{2}{15}]$ it is bound entangled,
which is detected by the realignment criterion~(\ref{realignment}) as well as by the later introduced MUB-witness~\cite{SHBAH2012-MUB,HMBRBCh2021_MUB,BBChHM2021_MUB}.

\section{Action of Lorentz boosts on quantum states}

Bound entanglement can be observed in a two-particle system with at least $3\otimes 3$
dimensions, where in turn the PPT criterion is only necessary but not sufficient for entanglement. Thus, to analyze the behavior of bound entanglement under Lorentz
boost, we consider here a system of two massive, relativistic spin-$1$ particles.
We identify the Hilbert space of states of such a particle,
$\mathcal{H}$, with the carrier space of the irreducible, unitary massive representation
of the Poincar\'e group for spin $1$.
The space $\mathcal{H}$ is spanned by the eigenvectors of the four-momentum operator
$\ket{k,\sigma}$, where $k=(k^0,\vec{k})$, $k^2={k^0}^2-\vec{k}^2=m^2$,
denotes the four-momentum of the particle and $\sigma=-1,0,1$ its spin component along
$z$-axis.

We denote here space-time coordinates with Greek indices running from $0$ to $3$, four-vectors
by plain letters, spacial vectors by bold letters, e.g.,
$k=(k^0,\vec{k})$. The Minkowski tensor is assumed to be
$\eta=\text{diag}(1,-1,-1,-1)$. We also use natural units with $c=\hbar=1$.

We use the Lorentz-covariant normalization
\begin{equation}
	\label{normalization_base}
	\bracket{k,\sigma}{k',\sigma'}=
	2k^0\delta^3\;(\vec{k}-\vec{k}^{\prime})\cdot\delta_{\sigma\sigma'}.
\end{equation}
The vectors $\ket{k,\sigma}$ can be generated from the standard vector
$\ket{\tilde{k},\sigma}$, where $\tilde{k}=m(1,0,0,0)$ is the
four-momentum of the particle in its rest frame. We have
$\ket{k,\sigma}=U(L_k)\ket{\tilde{k},\sigma}$, where the standard Lorentz boost
$L_k$ is defined by relations $k=L_k\tilde{k}$,
$L_{\tilde{k}}=\mathbbm{1}_4$. The explicit form of the boost $L_k$ reads
\begin{equation}
	L_k = \frac{1}{m}
	\begin{pmatrix}
		k^0 & \vec{k}^T \\
		\vec{k} & m\mathbbm{1}_3 + \frac{\vec{k}\otimes \vec{k}^T}{m+k^0}	
	\end{pmatrix}\;.
\end{equation}
With the help of the standard Wigner procedure~\cite{CRW2008,Caban2008,cab_BR1977} we get
\begin{equation}
	\label{U_Lambda_base}
	U(\Lambda)\ket{k,\sigma}=
	\mathcal{D}_{\lambda\sigma}(R(\Lambda,k))\ket{\Lambda	k,\lambda},
\end{equation}
where the Wigner rotation $R(\Lambda,k)$ is defined as
$R(\Lambda,k)=L_{\Lambda k}^{-1}\Lambda L_k$ and $\mathcal{D}$ is a
three dimensional, unitary, irreducible representation of the rotation group.
It is well known that in each dimension there exists, up to unitary equivalence,
only one unitary, irreducible representation of the rotation group.
Therefore, the representation $\mathcal{D}(R)$ is unitary equivalent
to $R$
\begin{equation}
	\label{equivalent}
	\mathcal{D}(R)=VRV^{\dag},\qquad V^\dagger V = \mathbbm{1}_3,
\end{equation}
and the explicit form of the matrix $V$ is the following:
\begin{equation}
	\label{matrix_V}
	V=\frac{1}{\sqrt{2}}
	\begin{pmatrix}
		-1 & i & 0 \\
		0 & 0 & \sqrt{2} \\
		1 & i & 0 \\
	\end{pmatrix}.
\end{equation}
For more details on spin-1 irreducible, unitary representation of the Poincare group
see, e.g., Refs.~\cite{CRW2008,Caban2008,cab_BR1977}.

For our computations we identify spin projection values $-1,0,1$ with indices
of computational basis vectors $0,1,2$ in the following way:
$-1 \leftrightarrow 0$,	$0 \leftrightarrow 1$, $1 \leftrightarrow 2$.

\section{Bound entanglement under Lorentz boosts}

Now, let us consider two inertial frames, $\mathcal{O}$ and $\mathcal{O}^\prime$,
and let the frame $\mathcal{O}^\prime$ move with the velocity $\vec{v}$
with respect to the frame $\mathcal{O}$.
In the frame $\mathcal{O}$ we prepare a two-particle state $\rho$.
For simplicity we treat momentum degrees of freedom as discrete, i.e.
we assume that momenta of the particles are chosen from the finite set
$\{k_1,\dots,k_N\}$. To prove all of our results it is enough to limit to
only two momenta $\{k_1,k_2\}$ with
\begin{equation}
k_{1,2}=(k^0,\pm |\vec{k}|,0,0)
\label{k1k2}
\end{equation}
Consequently, our total state under interest acts in
$(2\otimes 2)_{\mathsf{mom}}\otimes(3\otimes 3)_{\mathsf{spin}}=36$
dimensional space, i.e. in ${\mathbb{C}}^{36}$.
Of course, also higher dimensions in the momentum space are possible,
but not necessary to obtain our results.
Moreover, without loss of generality we can take
\begin{equation}
	m=1,\quad k^0 = 1+E,\quad |\vec{k}|=\sqrt{E(2+E)},
	\label{kinetic-energy}
\end{equation}
where $E$ is a kinetic energy of the particle.
Thus, the most general two-particle state we consider is of the form
\begin{equation}
\rho = \sum
\rho_{mn,m^\prime n^\prime}^{\sigma\lambda,\sigma^\prime \lambda^\prime}
\ket{k_m,k_n;\sigma,\lambda}
\bra{k_{m^\prime},k_{n^\prime};\sigma^\prime,\lambda^\prime},
\end{equation}
where $m,n,m^\prime,n^\prime=1,2$,
$\sigma,\lambda,\sigma^\prime,\lambda^\prime=0,1,2$ and
$\rho_{mn,m^\prime n^\prime}^{\sigma\lambda,\sigma^\prime \lambda^\prime}$
fulfill all necessary conditions to guarantee that $\rho$ is a valid density matrix.
We have also reordered the products of momentum and spin
components, i.e.
$\ket{k,p;\sigma,\lambda}=\ket{k,p}\otimes\ket{\sigma,\lambda}
\equiv \ket{k,\sigma}\otimes\ket{p,\lambda}$.

The state $\rho$ as seen from the frame ${\mathcal{O}}^\prime$ has
the following form
\begin{equation}
\rho^\prime = [U(\Lambda(\vec{e},\xi))\otimes U(\Lambda(\vec{e},\xi))]
\rho
[U(\Lambda(\vec{e},\xi))\otimes U(\Lambda(\vec{e},\xi))]^\dagger.
\end{equation}
Here $\Lambda(\vec{e},\xi)$ is the
Lorentz boost in the direction $\vec{e}=\vec{v}/|\vec{v}|$ with rapidity
$\xi$, $\tanh\xi=-|\vec{v}|$, joining frames $\mathcal{O}$ and $\mathcal{O}^{\prime}$.
Its explicit form is the following:
\begin{equation}
	\Lambda(\vec{e},\xi) =
	\begin{pmatrix}
		\cosh \xi & \vec{e}^T \sinh\xi \\
		\vec{e} \sinh\xi & \mathbbm{1}_3 + (\cosh\xi - 1)\vec{e}\otimes\vec{e}^T
	\end{pmatrix}.
	\label{boost-general}
\end{equation}
The action of $U(\Lambda)$ on basis states of the space ${\mathcal{H}}$ is given in
Eq.~(\ref{U_Lambda_base}).

The spin parts of the states $\rho$ and $\rho^{\prime}$
(i.e. $\rho_\mathsf{spin}$ and $\rho_{\mathsf{spin}}^{\prime}$, respectively)
we obtain by tracing out momentum degrees of freedom and normalizing the result
since in covariant normalization (\ref{normalization_base}) basis vectors
are orthogonal but not orthonormal.
Of course in the frame ${\mathcal{O}}^\prime$ momenta of the particles
belong to the set $\{k_{1}^\prime,k_{2}^\prime\}$, where
$k_{i}^\prime = \Lambda(\vec{e},\xi) k_i$, $i=1,2$.

\subsection{Pure momentum part of the state}

As the first case we consider the simple situation when
in the frame $\mathcal{O}$ we prepare a two-particle state
\begin{equation}
	\rho=\sum_i p_i \ket{\psi_i}\bra{\psi_i},\quad p_i\geq 0\quad\mathrm{with}\quad \sum_i p_i=1\;,
	\label{rho-initial}
\end{equation}
where
\begin{equation}
	\ket{\psi_i} = \ket{\psi^{\mathsf{mom}}}\otimes \ket{\varphi_i^\mathsf{spin}},
	\quad \ket{\psi_i}\in \mathcal{H}^\mathsf{mom}\otimes \mathcal{H}^\mathsf{spin},
	\label{psi_i}
\end{equation}
and $\ket{\psi^{\mathsf{mom}}}$, $\ket{\varphi_i^\mathsf{spin}}$ are momentum and spin parts
of the state $\ket{\psi_i}$, respectively.
Thus, we assume that the momentum parts of
all of the states $\ket{\psi_i}$ are identical.
Notice that the full state $\rho$ defined in Eqs.~(\ref{rho-initial},\ref{psi_i})
is separable with respect to the
partition: momenta versus spins (although it is not the most general separable
state, the general separable state we consider in the next section).
Therefore, we can write (\ref{rho-initial}) as
\begin{align}
\rho & = \ket{\psi^{\mathsf{mom}}}\bra{\psi^{\mathsf{mom}}}
\otimes \big( \sum_i p_i \ket{\varphi_i^\mathsf{spin}}\bra{\varphi_i^\mathsf{spin}} \big) \nonumber\\
& \equiv \ket{\psi^{\mathsf{mom}}}\bra{\psi^{\mathsf{mom}}}
\otimes \rho_{\mathsf{spin}},
\label{rho-pure-momentum}
\end{align}
and of course $\rho_{\mathsf{spin}} = \tr^{\mathsf{mom}}(\rho)$.
We are interested in the situation when $\rho_{\mathsf{spin}}$ ia a bound entangled
state.

The most general form of $\ket{\psi^{\mathsf{mom}}}$ in our case reads
\begin{equation}
\ket{\psi^{\mathsf{mom}}} = \sum_{i,j=1}^{2} a_{ij} \ket{k_i,k_j},
\quad \sum_{i,j=1}^{2} |a_{ij}|^2=1.
\end{equation}
Now, we boost the state (\ref{rho-pure-momentum}) and the spin part of the
boosted state has the form
\begin{multline}
\rho_{\mathsf{spin}}^\prime  = 	\tr^{\mathsf{mom}}(\rho^\prime)
 = \sum_{i,j=1}^{2} |a_{ij}|
\big[ \mathcal{D}^T(\Lambda,k_i) \otimes \mathcal{D}^T(\Lambda,k_j)  \big]\\
\rho_{\mathsf{spin}}
\big[ \mathcal{D}^*(\Lambda,k_i) \otimes \mathcal{D}^*(\Lambda,k_j)  \big].
\end{multline}
Following \cite{HHH_1998-bound} and using the property
\begin{equation}
\Big[ (A\otimes B)\rho (C\otimes D) \Big]^{T_B} =
(A\otimes D^T) \rho^{T_B} (C\otimes B^T)
\label{partial-tranposition-property}
\end{equation}
we easily see that if $\rho_{\mathsf{spin}}$ is a PPT state than also
$\rho_{\mathsf{spin}}^\prime$ is also PPT.
Thus, if the spin state of the two-particle state (\ref{rho-pure-momentum})
is bound entangled then for all other inertial observers this state is PPT, i.e.
also bound entangled or separable. We now show that both of these cases
can be realized. To this end we assume that
\begin{equation}
	\ket{\psi^{\mathsf{mom}}} =
	\ket{\psi^{\mathsf{mom}}_1} =
	\frac{1}{\sqrt{2}}
	(\ket{k_1,k_2} + \ket{k_2,k_1}).
	\label{momentum-state-chosen}
\end{equation}
where $k_1,k_2$ are given in Eq.~(\ref{k1k2}).
Moreover, as a spin part of the state (\ref{rho-pure-momentum}) we take
the state $\rho_b$ defined in Eq.~(\ref{rho-bound}).
Thus, the state chosen in the frame $\mathcal{O}$ has the form
\begin{equation}
\rho_1(x) =
\ket{\psi^{\mathsf{mom}}_1}\bra{\psi^{\mathsf{mom}}_1}
\otimes
\rho_b(x).
\label{rho_1}
\end{equation}
We further choose the
boost direction $\vec{e}=(0,0,1)$ (compare with Eq.~(\ref{boost-general})) and kinetic energy of a particle $E=1$ (compare with Eq.~(\ref{kinetic-energy})).
\begin{figure}
\includegraphics[width=1\columnwidth,keepaspectratio=true]{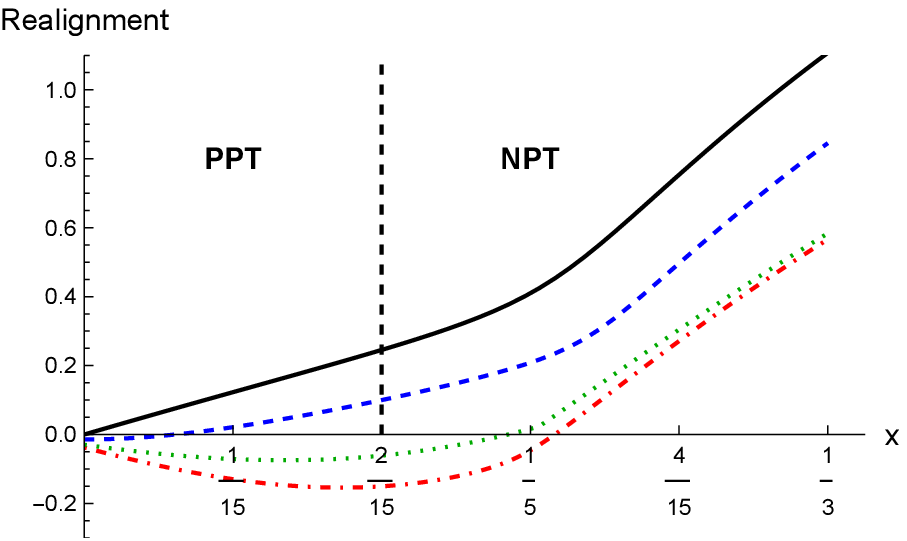}
\caption{Value of the base-2 log of the sum of all singular values of the realigned
spin density matrix for unboosted state (black, solid line) and states boosted
with rapidity $\xi=0.5$ (blue, dashed line), $\xi=0.8$ (green, dotted line),
$\xi=1$ (red, dashed-dotted line).}
\label{fig-realignment}
\end{figure}

In Fig.~\ref{fig-realignment} we plotted
the value of the base-2 log of the sum of all singular values of the realigned
spin density matrix versus $x$ for the unboosted and boosted state (\ref{rho_1}).
According to the realignment criterion~\cite{ChenWu2003-realignment},
if this value is positive the state is entangled.
We can see that for the unboosted state the realignment criterion
detects entanglement for all values of $x$ except $0$,
in contrast to the  boosted state $\rho_{\mathsf{spin}}^\prime(x,\xi)$ where
it detects entanglement only for $x\in(x_0,1/3]$.
Thus, in the considered case there always exist such a $x$ and $\xi$
that the bound entangled state $\rho_{\mathsf{spin}}=\rho_b(x)$
after boost is also bound entangled.

\begin{figure}
\includegraphics[width=1\columnwidth,keepaspectratio=true]{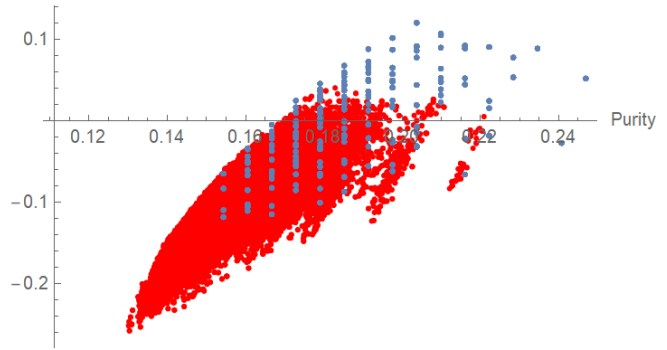}
\caption{This picture visualizes the purity $\tr\rho^2$ versus the realignment $\sum_i \sigma_i(\tilde{\rho}_{AB})-1$
of $156~600$ bound entangled magic simplex states~\cite{Hiesmayr2021}.
Those states result in $2~228$ (blue) points and if boosted with $\xi=0.8$ they
result in $87~116$ (red) points for our particular choice of boost.}
\label{fig-MagicSimplex}
\end{figure}

In Fig.~\ref{fig-MagicSimplex} we plotted the purity $\tr(\rho^2)$ versus realignment
of $156~600$ magic bound entangled states on a grid in the magic simplex~\cite{Hiesmayr2021},
where only some are detected by the realignment criterion. The other ones are detected by different criteria.
After the boost with $\xi=0.8$ the purity decreases and fewer states are still detected
by the realignment criterion. Thus the interesting question is:
Does a boost of a bound entangled state result sometimes also in separable states?

To find that out we can try to apply other entanglement criteria such as
the quasispin criterion~\cite{Hiesmayr2009-Quasispin} or the MUB
witness~\cite{SHBAH2012-MUB} which failed.
Thus, for e.g. $x=\frac{1}{15},\xi=0.8$ we tried to
show that $\rho_{\mathsf{spin}}^\prime$ is separable and were successful.
We performed this by numerically minimizing the Hilbert-Schmidt
distance $\big[\tr\big((\sum_{i=1}^{k} p_i^\mathsf{sep} \ket{\psi^\mathsf{sep}_i}\bra{\psi^\mathsf{sep}_i}
-\rho_{\mathsf{spin}}^\prime(\tfrac{1}{15},\tfrac{4}{5}))^2\big)\big]^{1/2}$, where $p_i^\mathsf{sep}>0$,
$\sum_{i=1}^{k} p_i^\mathsf{sep}=1$, and
$\ket{\psi^\mathsf{sep}_i}$ represent separable states,
for some large enough $k$.
For $k=10$ we found
probabilities $p_i^\mathsf{sep}$ and vectors $\ket{\psi^\mathsf{sep}_i}$ such that
\begin{equation}
\Big[\tr\big((\sum_{i=1}^{10} p_i^\mathsf{sep} \ket{\psi^\mathsf{sep}_i}\bra{\psi^\mathsf{sep}_i}
-\rho_{\mathsf{spin}}^\prime(\tfrac{1}{15},\tfrac{4}{5}))^2\big)\Big]^{1/2} \approx 7\times 10^{-8},
\end{equation}
i.e. practically equal to zero.
The explicit form of probabilities $p_i^\mathsf{sep}$ and vectors
$\ket{\psi^\mathsf{sep}_i}$ we give in Appendix \ref{app:A}. We were also successful for other $\xi$ values and other $x$ values like $x=\frac{1}{10}$, but not for $x>\frac{2}{15}$, the NPT area.

In summary, we have found at least some states which are after the boost separable,
which is apparently against the intuition if it depends on the boost of the observer
whether a state is classified as bound entangled or separable.

\subsection{General separable state momenta versus spins}

Now, let us assume that in the frame $\mathcal{O}$ we prepare the most general
state that is separable in the partition momenta versus spins, i.e.
\begin{equation}
\rho = \sum_i p_i\;  \rho_{\mathsf{mom}}^i \otimes \rho_{\mathsf{spin}}^i,
\label{rho_sep_general}
\end{equation}
where $p_i>0$, $\sum_i p_i=1$. The spin part of this state has the form
\begin{equation}
\rho_{\mathsf{spin}} = \tr_{\mathsf{mom}}(\rho)
= \sum_i p_i\;  \rho_{\mathsf{spin}}^i.
\label{rho-spin-general-separable}
\end{equation}
As previously, we are interested in the situation when the whole state
$\rho_{\mathsf{spin}}$ is bound entangled.

Now, we boost the state (\ref{rho_sep_general}) and calculate the spin part of the
boosted state as $\rho_{\mathsf{spin}}^\prime = \tr^{\mathsf{mom}}(\rho^\prime)$
and we obtain
\begin{multline}
\rho_{\mathsf{spin}}^\prime  =
	\sum_{i,m,n} p_i c_{mn,mn}^i\;
	\big[ \mathcal{D}^T(\Lambda,k_m) \otimes \mathcal{D}^T(\Lambda,k_n)  \big]
	\rho_{\mathsf{spin}}^i \\
	\big[ \mathcal{D}^*(\Lambda,k_m) \otimes \mathcal{D}^*(\Lambda,k_n)  \big],
\label{rho-spin-prime-general-separable}
\end{multline}
where
\begin{equation}
\rho_{\mathsf{mom}}^i =
\sum_{m,n,m^\prime,n^\prime}
c_{mn,m^\prime n^\prime}^i\;
\ket{k_m,k_n}\bra{k_{m^\prime},k_{n^\prime}},
\end{equation}
i.e. $c_{mn,mn}^i>0$, $\sum_{m,n} c_{mn,mn}^i =1$.

Now, we ask whether the transformation (\ref{rho-spin-prime-general-separable}), the relativistic boost,
preserves PPT. However, applying Eq.~(\ref{partial-tranposition-property})
we have to take into account that the whole state $\sum_i p_i  \rho_{\mathsf{spin}}^i$
is PPT but some of the states $\rho_{\mathsf{spin}}^i$ can be NPT. Next,
factors $c_{mn,mn}^i$ change relative weights in the sum
(\ref{rho-spin-prime-general-separable}). Thus, we cannot conclude that
(\ref{rho-spin-prime-general-separable}) preserves PPT. In fact, we can find such states
and boosts that (\ref{rho-spin-general-separable}) is bound entangled while
(\ref{rho-spin-prime-general-separable}) is free entangled.
For example, let us consider the following state:
\begin{equation}
\rho_0(p,x) = p\; \rho_1(x) + (1-p)\; \rho_2,
\end{equation}
where $0\le p \le 1$, $\rho_1(x)$ is given in Eq.~(\ref{rho_1}) and
\begin{equation}
\rho_2 = \ket{k_1,k_2}\bra{k_1,k_2} \otimes
\ket{\varphi_\mathsf{spin}}\bra{\varphi_\mathsf{spin}},
\end{equation}
with
\begin{equation}
	\ket{\varphi_\mathsf{spin}} = \sum_{k,l=0}^{2} a_{k,l} \ket{\Omega_{k,l}},
	\label{rho-spin-2}
\end{equation}
and
\begin{equation}
	\begin{pmatrix}
		a_{0,0}& a_{0,1}& a_{0,2}\\
		a_{1,0}& a_{1,1}& a_{1,2}\\
		a_{2,0}& a_{2,1}& a_{2,2}
	\end{pmatrix}
	=
	\begin{pmatrix}
		0& 2/9& 2/9\\
		0& 2/9& 1/18\\
		5/18& 0& 0
	\end{pmatrix}.
\end{equation}
The spin part of the state $\rho_0(0.04,\tfrac{7}{60})$,
$\rho_0^{\mathsf{spin}}(0.04,\tfrac{7}{60})$ is PPT and
\begin{equation}
	\RLGMT(\rho_0^{\mathsf{spin}}(0.04,\tfrac{7}{60})) = 0.183 >0,
\end{equation}
thus it is bound entangled.

Now, if we boost this state in the direction $\vec{e}=(0,0,1)$ with $\xi=0.95$
we obtain the state $\rho_0^\prime(0.04,\tfrac{7}{60})$.
The spin part $\rho_0^{\prime,\mathsf{spin}}(0.04,\tfrac{7}{60})$ is not PPT, thus it is free
entangled and obviously no longer bound entangled.

In summary, we have shown that starting from a bound entangled state, we can boost
to a separable, bound entangled or an free entangled state.
In the next section we discuss how those counter-intuitive classifications of inertial
observers relates to the principle that the physics observed, i.e. probabilities and
expectation values, should be Lorentz invariant.

\section{Physics discussion}
\label{MUBWitness}

Obviously, the physics for every observer should be the same.
In our case this means that the values that a boosted or not boosted observer obtains
by computing $\tr(\mathcal{O}\rho)$ for some observable $\mathcal{O}=\mathcal{O}^\dagger$ are identical.
An observable $\mathcal{W}$ for which
\begin{equation}
\min_{\rho_\mathsf{sep}}(\tr(\mathcal{W}\rho_\mathsf{sep}))
\leq \tr(\mathcal{W}\rho)
\leq \max_{\rho_\mathsf{sep}}(\tr(\mathcal{W}\rho_\mathsf{sep}))
\end{equation}
does not hold for all states $\rho$ is called an entanglement witness.
The upper and lower bounds define the so called separability window of the witness
and can be obtained by mirroring the witness~\cite{BCH_2020_mirorred-witnesses}.
A particular witness, decomposable into mutually unbiased bases (MUBs), is capable
to detect bound entanglement~\cite{SHBAH2012-MUB,HMBRBCh2021_MUB,BBChHM2021_MUB}
and as it contains a recipe how to realize it experimentally, it gave raise to
the first experiment detecting bound entanglement in bipartite systems~\cite{HL2013}.
We will use this witness to show how we can solve the apparent puzzle. Obviously, we have
\begin{equation}
 \tr(\mathcal{W}\rho)=
 \tr(\mathcal{W}^{\mathsf{boosted}}\rho^{\mathsf{boosted}})
 \label{equal}
\end{equation}
for any chosen boost on the \textit{total} space. But this does not mean that this is
the case also for the subsystem, i.e. an partition into first\&second momentum versus
first\&second spin or first momentum\&first spin
versus second momentum\;second momentum nor
first momentum\& second spin versus second momentum\&first spin nor in particle
A (first momentum/first spin) versus particle B (second momentum/first spin)
as we show in the following (for more details on the behavior of entanglement
under different partitions see, e.g., Ref.~\cite{FBHH2010}).

The MUB-witness for two qudits that is capable of detecting bound entanglement
by the upper bound is defined by
\begin{eqnarray}
\tr(\mathcal{W}\rho)&=&\sum_{i=0}^{d-1} \bra{i_1}\otimes \bra{(i_1+1)^*} \rho
  \ket{i_1}\otimes \ket{(i_1+1)^*}\nonumber\\
  &&+\sum_{k=2}^{d+1}\sum_{i=0}^{d-1} \bra{i_k}
  \otimes \bra{i_k^*} \rho \ket{i_k}
  \otimes
  \ket{i_k^*}\;,
\end{eqnarray}
for which the orthonormal basis $\{i_k\}_{k=1}^{d+1}$ are mutually unbiased
$|\langle i_k|j_l\rangle|^2=\delta_{k,l} \delta_{i,j}+(1-\delta_{k,l})\frac{1}{d}$.
Thus we can boost each MUB vector and by that achieve the observable
$\mathcal{W}^{\mathsf{boosted}}$ that is seen by a boosted observer.
On the total space we find for our particular example
(optimization of the lower and upper bound via the composite
parameterization~\cite{SHH2012})
\begin{eqnarray}
-\frac{3}{4}\leq \tr(\mathcal{W}\rho)
= \tr(\mathcal{W}^{\mathsf{boosted}}\rho^{\mathsf{boosted}})
=\frac{1}{2}+\frac{x}{4}\leq \frac{3}{4}\nonumber\\
\end{eqnarray}
for every boost value $\xi$. This inequality is obviously
not violated since our total state was chosen to be separable (momentum versus spin).

Obviously, the situation changes when we consider the spin or momentum subspaces,
which we achieve by partial tracing over the respective subsystems.
For the spin part we find for the unboosted case
\begin{eqnarray}
\frac{2}{3}\;\leq\; \tr(\tr_\mathsf{mom} W \tr_\mathsf{mom} \rho)=2+x\;\leq 2
\end{eqnarray}
which is obviously violated if $x\not=0$. And e.g. for $\xi=0.8$ we have
\begin{eqnarray}
0.763&\leq& \tr(\tr_\mathsf{mom} W^{\mathsf{boosted}} \tr_\mathsf{mom} \rho^{\mathsf{boosted}})\nonumber\\
&&=1.694 + 0.641 x\;\leq\;1.985\;,
\end{eqnarray}
i.e. no violation for any $x$.
Consequently, if one ignores in the boosted case the momentum degrees of freedom,
one does not consider the full experimental situation and this explains why
an unboosted bound state becomes a separable, bound entangled or free entangled
one when ignoring part of the systems, though, against our intuition,
the momentum part of the witness is just the unity operator for any boost.

\section{Summary}

In this paper we analyzed the behavior of the bound entangled states under Lorentz
boosts. To this end we considered a system of two spin-1, massive particles.
In a given inertial frame of reference this system is prepared in a state which
is separable with respect to the partition momenta/spins and such that its spin
part is bound entangled. Next, we boost this state to a different inertial frame
of reference and analyze the entanglement of its spin part.
We showed that the boosted state can be also bound entangled.
However, we were also able to find such states and boosts that
the boosted state is separable or even free entangled.
Thus, surprisingly, we found that Lorentz boosts can activate bound entanglement.

In the last section we explained how such a counter-intuitive classification
of different inertial observers relates to the principle that measurement outcomes
are invariant under Lorentz boosts.
Boosting a state means that one boosts also the observable which means
taking the trace gives the same result. But ignoring the momentum part,
the situation changes drastically and leads to all three separability possibilities
as we have shown by exemplary boosts and states.

From the purely  practical point of view, we have presented a method to systematically
produce different bound entangled states by application of different relativistic boosts,
which may further be explored to reveal the very Nature of bound entanglement.

\vspace{0.5cm}

\textbf{Acknowledgments:} PC is supported by University of Lodz under the IDUB project. BCH acknowledges gratefully that this research was funded in whole, or in part, by the   Austrian Science Fund (FWF) project P36102-N. For the purpose of open access, the author has applied a CC BY public copyright license to any Author Accepted Manuscript version arising from this submission.

\appendix
\section{Separable state}
\label{app:A}
Below we give an explicit form of probabilities $p_i^\mathsf{sep}$ and states
$\ket{\psi^\mathsf{sep}_i}$ for which
\begin{equation}
	\Big[\tr\big((\sum_{i=1}^{10} p_i^\mathsf{sep} \ket{\psi^\mathsf{sep}_i}\bra{\psi^\mathsf{sep}_i}
	-\rho_{\mathsf{spin}}^\prime(\tfrac{1}{15},\tfrac{4}{5}))^2\big)\Big]^{1/2}
	\approx 7\times 10^{-8},
\end{equation}
where $\rho_{\mathsf{spin}}^\prime(\tfrac{1}{15},\tfrac{4}{5})$
is a spin part of a boosted state (\ref{rho_1}) for $x=\tfrac{1}{15}$,
$\xi=\tfrac{4}{5}$.
\begin{widetext}
\begin{multline}
\{p_1^\mathsf{sep},p_2^\mathsf{sep},\dots,p^\mathsf{sep}_{10}\} =
\{0.1430992, 0.0734831,
0.0852185, 0.1018063, 0.1491225, 0.0693471,
0.0776687, 0.1398679,\\
0.0991522, 0.0612343\}
\end{multline}
\begin{equation}
\psi_1^\mathsf{sep} =
\begin{pmatrix}
-0.0430823\\
0.0167110 + 0.0537862 i\\
-0.8497352 - 0.5175507 i\\
-0.0005030 - 0.0027742 i\\
-0.0032683 + 0.0017040 i\\
0.0234053 - 0.0607592 i\\
0.0012085 - 0.0001888 i\\
-0.0007044 - 0.0014356 i\\
0.0261032 + 0.0107947 i
\end{pmatrix},
\psi_2^\mathsf{sep} =
\begin{pmatrix}
	-0.1680125\\
	-0.0326983 - 0.4851121 i\\
	0.1438131 - 0.5714999 i\\
	-0.0775247 + 0.0557386 i\\
	-0.1760249 - 0.2129939 i\\
	-0.1232379 - 0.3114132 i\\
	-0.0481015 + 0.0804443 i\\
	-0.2416331 - 0.1232304 i\\
	-0.2324608 - 0.2324767 i
\end{pmatrix},
\psi_3^\mathsf{sep} =
\begin{pmatrix}
	0.2547478\\
	0.2422784 - 0.0536082 i\\
	-0.1178008 + 0.0076515 i\\
	0.4163048 + 0.0276915 i\\
	0.4017548 - 0.0612697 i\\
	-0.1933399 - 0.0003012 i\\
	-0.4580805 - 0.1157550 i\\
	-0.4600175 - 0.0136921 i\\
	0.2153030 + 0.0397689 i
\end{pmatrix}
\end{equation}
\begin{equation}
\psi_4^\mathsf{sep} =
\begin{pmatrix}
	0.2077491\\
	-0.2713063 + 0.3203387 i\\
	0.5379288 - 0.2613437 i\\
	-0.1363512 - 0.0707598 i\\
	0.2871735 - 0.1178391 i\\
	-0.4420709 - 0.0116930 i\\
	0.0892001 - 0.0088623 i\\
	-0.1028240 + 0.1491157 i\\
	0.2198189 - 0.1351591 i
\end{pmatrix},
\psi_5^\mathsf{sep} =
\begin{pmatrix}
	-0.0015317\\
	0.0236887 + 0.0010515 i\\
	0.0009902 - 0.0043033 i\\
	0.0021656 - 0.0073004 i\\
	-0.0385048 + 0.1114204 i\\
	0.0191106 + 0.0108040 i\\
	0.0626658 - 0.0053981 i\\
	-0.9728846 + 0.0404651 i\\
	-0.0253473 + 0.1795503 i
\end{pmatrix},
\psi_6^\mathsf{sep} =
\begin{pmatrix}
	-0.3732019\\
	-0.2337885 + 0.1894137 i\\
	-0.3174714 + 0.3648181 i\\
	-0.2753247 - 0.1903291 i\\
	-0.2690733 + 0.0205078 i\\
	-0.4202637 + 0.1072325 i\\
	-0.0215085 - 0.2204501 i\\
	-0.1253603 - 0.1271824 i\\
	-0.2337944 - 0.1665049 i
\end{pmatrix},
\end{equation}
\begin{equation}
\psi_7^\mathsf{sep} =
\begin{pmatrix}
	0.0437898\\
	-0.0790503 + 0.1006166 i\\
	-0.0675530 - 0.0621037 i\\
	-0.1008128 + 0.0029161 i\\
	0.1752890 - 0.2369033 i\\
	0.1596559 + 0.1384763 i\\
	-0.0287347 + 0.2426421 i\\
	-0.5056504 - 0.5040468 i\\
	0.3884482 - 0.3335631 i
\end{pmatrix},
\psi_8^\mathsf{sep} =
\begin{pmatrix}
	-0.2327074\\
	-0.0014722 - 0.0295408 i\\
	0.0033542 - 0.0039737 i\\
	-0.3230604 + 0.9062410 i\\
	-0.1170858 - 0.0352773 i\\
	-0.0108184 - 0.0185789 i\\
	-0.0454358 + 0.0419600 i\\
	-0.0056140 - 0.0055023 i\\
	-0.0000616 - 0.0013807 i
\end{pmatrix},
\psi_9^\mathsf{sep} =
\begin{pmatrix}
	0.4046742\\
	-0.1996496 - 0.1249451 i\\
	-0.1091446 + 0.0005520 i\\
	-0.5066867 + 0.3689981 i\\
	0.3639082 - 0.0256066 i\\
	0.1361550 - 0.1002136 i\\
	-0.3127123 + 0.2324570 i\\
	0.2260516 - 0.0181333 i\\
	0.0840245 - 0.0631225 i
\end{pmatrix},
\end{equation}
\begin{equation}
\psi_{10}^\mathsf{sep} =
\begin{pmatrix}
	-0.3632900\\
	-0.0910701 + 0.0798075 i\\
	-0.0632746 - 0.1351573 i\\
	0.1483556 - 0.5829032 i\\
	-0.0908620 - 0.1787139 i\\
	0.2427008 - 0.0463310 i\\
	-0.2349668 + 0.4820461 i\\
	0.0469939 + 0.1724576 i\\
	-0.2202634 - 0.0034579 i
\end{pmatrix}.
\end{equation}
\end{widetext}


%

\end{document}